\documentclass[12pt]{article}

\date{}
\begin{document}

\title{\bf Shannon Entropy: Axiomatic Characterization and Application}
\author{C. G. Chakrabarti,Indranil Chakrabarty}
\vskip .1in

\begin{titlepage}

\maketitle

\noindent

\vspace{7em}

We have presented a new axiomatic derivation of Shannon Entropy
for a discrete probability distribution on the basis of the
postulates of additivity and concavity of the entropy function.We
have then modified shannon entropy to take account of
observational uncertainty.The modified entropy reduces, in the
limiting case, to the form of Shannon differential entropy.As an
application we have derived the expression for classical entropy
of statistical mechanics from the quantized form of the entropy.\\


\vspace{3em}

\end{titlepage}

\setcounter{page}{2}

\baselineskip .3in

\noindent {\large\bf 1. Introduction}

\vspace{1em}

Shannon entropy is the key concept of information theory [1]. It
has found wide applications in different fields of science and
technology [2-5]. It is a characteristic of probability
distribution providing a measure of uncertainty associated with
the probability distribution. There are different approaches to
the derivation of Shannon entropy based on different postulates or
axioms [6, 7].\\

The object of present paper is to stress the importance of the
properties of additivity and concavity in the determination of
functional form of Shannon entropy and it's generalization. The
main content of the paper is divided into three sections. In
section 2 we have provided an axiomatic derivation of Shannon
entropy on the basis of the properties of additivity and concavity
of entropy-function. In section 3 we have generalized Shannon
entropy and introduced the notion of total entropy to take account
of observational uncertainty. The entropy of continuous
distribution, called the differential entropy has been obtained as
a limiting value . In section 4 the differential entropy along
with the quantum uncertainty relation has been used to derive the
expression of classical entropy in statistical mechanics.

\vspace{2em}

\noindent{\large\bf 2. Shannon Entropy : Axiomatic
Characterization}

\vspace{1em}

Let $\Delta_{n}$ be the set of all finite discrete probability
distribution $$P = \{(p_{1},p_{2},....,p_{n}), p_{i}\geq 0,
\sum_{i=1}^{n}p_{i}=1\}$$ In other words, $P$ may be considered
as a random experiment having n possible outcomes with
probabilities $(p_{1},p_{2},....,p_{n})$. There is uncertainty
associated with the probability distribution $P$ and there are
different measures of uncertainty depending on different
postulates or conditions. In general, the uncertainty associated
with the random experiment $P$ is a mapping [8]
\begin{center}$$H(P) :\Delta_{n}\rightarrow
\mathbf{R}\eqno(2.1)$$\end{center} where $\mathbf{R}$ is the set
of real numbers. It can be shown that (2.1) is a reasonable
measure of uncertainty if and only if it is a Shur concave on
$\Delta_{n}$ [8]. A general class of uncertainty measures is given
by
\begin{center}$$ H(p) = \sum_{i=1}^{n}\phi(p_{i})
\eqno(2.2)$$\end{center} where $\phi : [0,1]\rightarrow
\mathbf{R}$ is a concave function. By taking different concave
function defined on [0,1], we get different measures of
uncertainty or entropy. For example, if we take $\phi(p_{i}) =
-p_{i}\log p_{i}$, we get Shannon entropy [1] \begin{center}$$
H(P) = H(p_{1},p_{2},....,p_{n}) = -k\sum_{i=1}^{n}p_{i}\log p_{i}
\eqno(2.3)$$\end{center} where $0 \log 0 = 0$ by convention and
$k$ is a constant depending on the unit of measurement of entropy.
There are different axiomatic characterizations of Shannon entropy
based on different set of axioms [6,7]. In the following we shall
present a different approach depending on the concavity character
of entropy-function. We set the following axiom to be satisfied
by the entropy function $ H(P) =H(p_{1},p_{2},....,p_{n})$.\\\

{\bf Axiom (1)} : We assume that the entropy $H(P)$ is
non-negative, that is, for all $P = (p_{1},p_{2},....,p_{n})$,
$H(P) \geq 0$. This is essential for a measure.\\\

{\bf Axiom (2)}: We assume that generalized form of
entropy-function (2.2):
\begin{center}$$H(P) = \sum_{i=1}^{n}\phi(p_{i}) \eqno(2.4)$$\\\ \end{center}

{\bf Axiom (3)} : We assume that the function $\phi$ is a
continuous concave
function of its arguments. \\\

{\bf Axiom (4)} : We assume the additivity of entropy, that is,
for any two statistically independent experiment $P =
(p_{1},p_{2},....p_{n})$ and $Q = (q_{1},q_{2},....,q_{m})$
\begin{center}$$H(PQ) = \sum_{j}\sum_{\alpha} \phi(p_{j} q_{\alpha}) =
\sum_{j}\phi(p_{j}) + \sum_{\alpha} \phi(q_{\alpha})
\eqno(2.5)$$\end{center} Then we have the following
theorem.\\\

{\bf THEOREM (2.1)} : If the entropy-function $H(P)$ satisfies the
above axioms (1) to (4),then $H(P)$ is given by\begin{center}$$
H(P)=-k\sum_{i=1}^{n}p_{i}\log p_{i}\eqno(2.6)$$\end{center} where
k is a positive constant depending on the unit of measurement of
entropy.\\\

{\bf PROOF} : For two statistically independent experiments the
joint probability distribution $p_{j\alpha}$ is the direct product
of the individual probability distributions
\begin{center}$$p_{j\alpha}=p_{j}.q_{\alpha} \eqno(2.7)$$\end{center}Then according to
the axiom of additivity of entropy (2.5), we
have\begin{center}$$\sum_{j}\sum_{\alpha}\phi(p_{j}.q_{\alpha})=\sum_{j}\phi(p_{j})+\sum_{\alpha}\phi(q_{\alpha})
\eqno(2.8)$$\end{center} Let us now make small changes of the
probabilities $p_{k}$ and $p_{j}$ of the probability distribution
$P = (p_{1},p_{2},....,p_{j},..p_{k},...,p_{n})$ leaving others
undisturbed and keeping the normalization condition fixed. By the
axiom of continuity of $\phi$ the relation (2.8) can be reduced
to the form \begin{center}
$$\sum_{\alpha}q_{\alpha}[\phi'(p_{j}.q_{\alpha})-\phi'(p_{k}.q_{\alpha}]=
\{\phi'(p_{j})-\phi'(p_{k})\}\eqno(2.9)$$\end{center} The r.h.s
of (2.9) is independent of $q_{\alpha}$and the relation (2.9) is
satisfied independently of $p$'s if \begin{center}
$$\phi'(q_{\alpha}.p_{j})-\phi'(q_{\alpha}p_{k}) =
\phi'(p_{j})-\phi'(p_{k})\eqno(2.10)$$ \end{center} The above
leads to the Cauchy's functional equation
\begin{center}$$\phi'(q_{\alpha}.p_{j}) = \phi'(q_{\alpha}) + \phi'(p_{j})
\eqno(2.11)$$\end{center} The solution of the functional equation
(2.11) is given by\begin{center} $$ \phi'(p_{j}) = A\log p_{j} + B
\eqno(2.12)$$\end{center} or \begin{center}$$\phi(p_{j}) =
Ap_{j}\log p_{j} + (B-A)p_{j} +C \eqno(2.13)$$\end{center} where
$A, B and C $are all constants. The condition of concavity
(axiom(3)) requires $A < 0$ and let us take $A = -k$ where $k (>
0)$ is positive constant by axiom (1). The generalized entropy
(2.4) then reduces to the form \begin{center}$$H(P) =
-k\sum_{j}p_{j}\log p_{j} + (B-A) + C \eqno(2.14)$$\end{center} or
\begin{center}$$ H(P) = -k\sum_{j}p_{j}\log p_{j} \eqno(2.15)$$\end{center} where
constants (B-A) and C have been omitted without changing the
character of the entropy-function. This proves the theorem.

\vspace{2em}

\noindent{\large\bf 3. Total Shannon Entropy and Entropy of
Continuous Distribution}

\vspace{1em}

The definition (2.3) of entropy can be generalized
straightforwardly to define the entropy of a discrete random
variable.\\\

{\bf DEFINITION} : Let $X \in \mathcal{R}$ denotes a discrete
random variable which takes on the values
$x_{1},x_{2},....,x_{n}$ with probabilities
$p_{1},p_{2},....,p_{n}$ respectively, the entropy $H(X)$ of $X$
is then defined by the expression [3] \begin{center}$$ H(X) =
-k\sum_{i=1}^{n}p_{i}\log p_{i} \eqno(3.1)$$\end{center}

 \vspace{1em}

 Let us now generalize the above definition to take account for an additional
uncertainty due to the observer himself, irrespective of the
definition of random experiment. Let $X$ denotes a discrete
random variable which takes the values $x_{1},x_{2},....,x_{n}$
with probabilities $p_{1},p_{2},....,p_{n}$. We decompose the
practical observation of $X$ into two stages. First, we assume
that $X \in L(x_{i})$ with probability $p_{i}$, where $L(x_{i})$
denotes the $i$th interval of the set
$\{L(x_{1}),L(x_{2}),....,L(x_{n})\}$ of intervals indexed by
$x_{i}$. The Shannon entropy of this experiment is $H(X)$.
Second, given that $X$ is known to be in the $i$th interval, we
determine its exact position in $L(x_{i})$ and we assume that the
entropy of this experiment is $U(x_{i})$. Then The global entropy
associated with the random variable $X$ is given by
\begin{center}$$ H_{T}(X) = H(X) + \sum_{i=1}^{n}p_{i}U(x_{i})
\eqno(3.2)$$\end{center} Let $h_{i}$ denotes the length of the
$i$th interval $L(x_{i})$, $( i=1,2,...,n )$, and define
\begin{center}$$ U(x_{i}) = k\log h_{i} \eqno(3.3)$$\end{center}
We have then \begin{center}$$H_{T}(X) = H(X) +
k\sum_{i=1}^{n}p_{i}\log h_{i} = -k\sum_{i=1}^{n}p_{i}\log
\frac{p_{i}}{h_{i}} \eqno(3.4)$$\end{center} The expression
$H_{T}(X)$ given by (3.4) will be referred to as the total entropy
of the random variable $X$. The above derivation is physical. In
fact, what we have used is merely a randomization of the
individual event $X = x_{i}$, $( i=1,2,....,n)$ to account for the
additional uncertainty due to the observer himself, irrespective
of the definition of random experiment [3]. We shall, derive the
expression (3.4) axiomatically as
generalization of the theorem (2.1).\\\

{\bf THEOREM (3.1)} : Let the generalized entropy (2.2)
satisfies, in addition to the axioms (1) to (4) of theorem (2.1)
the boundary conditions : \begin{center}$$ \phi_{i}(1) = k \log
h_{i}, ( i = 1,2,....,n) \eqno(3.5)$$\end{center} to take account
of the post-observational uncertainty where $h_{i}$ is the length
of the $i$th class $L(x_{i})$ ( or width of the observational
value $x_{i}$). Then the entropy-function reduces to the form of
the
total entropy (3.4).\\\

{\bf PROOF} : The procedure is the same as that of theorem (2.1)
upto the relation (2.12) : \begin{center}$$ \phi'(p_{j}) = A\log
p_{j} + B \eqno(3.6)$$\end{center} Integrating (3.6) with respect
to $p_{j}$and using the boundary condition (3.5), we have
\begin{center}$$\phi(p_{j})-k\log h_{j} = Ap_{j}\log p_{j} + (B-A)p_{j} - B
\eqno(3.7)$$\end{center} so that the generalized entropy (2.2)
reduces to the form \begin{center}$$ \sum_{j}\phi(p_{j}) =
-k\sum_{j = 1}^{n}p_{j}\log \frac{p_{j}}{h_{j}}
\eqno(3.8)$$\end{center} where we have taken $A = -k < 0$ for the
same unit of measurement of entropy and the negative sign to take
account the axiom (1). The constants appearing in (3.8) have been
neglected without any loss of characteristic properties. The
expression (3.8) is the required expression of
total entropy obtained earlier.\\
Let us now see how to obtain the entropy of a continuous
probability distribution as a limiting value of the total entropy
$H_{T}(X)$ defined above. For this let us first define the
differential entropy $H(X)$ of a continuous random variable
$X$.\\\

{\bf DEFINITION } : The differential entropy $H_{C}(X)$ of a
continuous random variable with probability density $f(x)$ is
defined by [9] \begin{center}$$ H_{C}(X) = -k \int_{R}f(x)\log
f(x) dx \eqno(3.9)$$\end{center} where $R$ is the support set of
the random variable $X$. We divide the range of $X$ into bins of
length ( or width ) h. Let us assume that the density $f(x)$ is
continuous within the bins. Then by mean value theorem, there
exists a value $x_{i}$ within each bin such that \begin{center}$$
hf(x_{i}) = \int_{ih}^{(i+1)h} f(x) dx \eqno(3.10)$$\end{center}
We define the quantized or discrete probability distribution $(
p_{1},p_{2},.....,p_{n})$ by \begin{center}$$ p_{i} =
\int_{ih}^{(i+1)h} f(x) dx \eqno(3.11)$$\end{center} so that we
have then \begin{center}$$ p_{i} = hf(x_{i}) \eqno(3.12)$$
\end{center} The total entropy $ H_{T}(X)$ defined for $h_{i} = h
( i = 1,2,....,n)$ \begin{center}$$H_{T}(X) =
-k\sum_{i=1}^{n}p_{i}\log \frac{p_{i}}{h_{}}
\eqno(3.13)$$\end{center} then reduces to the form
\begin{center}$$H_{T}(X) = -k\sum_{i=1}^{n}hf(x_{i})\log f(x_{i})
\eqno(3.14)$$\end{center} Let $ h \rightarrow 0$, then by
definition of Riemann integral we have $ H_{T}(X) \rightarrow
H(X)$ as $ h \rightarrow 0$, that is, \begin{center}$$ \lim_{h
\rightarrow 0}H_{T}(X) = H_{C}(X) = -k\int_{R} f(x) \log f(x) dx
\eqno(3.15)$$\end{center} Thus we
have the following theorem : \\\

{\bf THEOREM (3.2)} : The total entropy $H_{T}(X)$ defined by
(3.13) approaches to the differential entropy $H_{C}(X)$ in the
limiting case when the length of each bin tends to zero.

\vspace{2em}

\noindent{\large\bf 4.  Application:Differential Entropy and
Entropy in Classical Statistics  }

\vspace{1em}

The above analysis leads to an important relation connecting
quantized entropy and differential entropy. From (3.13) and
(3.15) we see that \begin{center}$$ -k \sum_{i=1}^{n} p_{i}\ln
p_{i} \rightarrow -k\int_{R} f(x) \ln \{hf(x)\} dx
\eqno(4.1)$$\end{center} showing that when $h \rightarrow 0$ that
is, when the length of the bins $h$ is very small the quantized
entropy given by the l.h.s of (4.1) approaches not to the
differential entropy $H_{C}(X)$ defined in (3.9) but to the form
given by the r.h.s of (4.1) which we call modified differential
entropy. This relation has important physical significance in
statistical mechanics. As an application of this relation we now
find the expression of
classical entropy as a limiting case of quantized entropy.\\

Let us consider an isolated system with configuration space volume
$V$ and a fixed number of particles $N$, which is constrained to
the energy-shell $R=( E, E+\Delta E )$. We consider the energy
shell rather than just the energy surface because the Heisenburg
uncertainty principle tells us that we can never determine the
energy $E$ exactly. we can make $\Delta E$ as small as we like.
Let $f(X^{N})$ be the probability density of microstates defined
on the phase space $ \Gamma = \{X^{N} = ( q_{1},q_{2},....,q_{2N};
p_{1},p_{2},....,p_{2N} )$ . The normalized condition is
\begin{center}$$ \int_{R} f(X^{N})X^{N} =1
\eqno(4.2)$$\end{center} where \begin{center}$$ R = \{ X^{N} : E <
H(X^{N}) < E + \Delta E \} \eqno(4.3)$$\end{center} Following
(4.1) we define the entropy of the system as \begin{center}$$ S =
-k \int f(X^{N}) \ln \{ C^{N} f(X^{N})\} dX^{N}
\eqno(4.4)$$\end{center} The constant $C^{N}$ appearing in (4.4)
is to be determined later on. The probability density for
statistical equilibrium determined by maximizing the entropy(4.4)
subject to the condition (4.2) leads to
\begin{center}$$f(X^{N}) = \frac{1}{\Omega( E, V, N )} \hspace{3em}
for\,\, E < H(X^{N}) < E + \Delta E
\eqno(4.5)$$\end{center} \hspace{12.4em}$=  0$ \hspace{7em} otherwise\\
 where $H(X^{N})$ is the Hamiltonian of the system, $\Omega (E,V,N)$ is the volume of
the energy shell $( E, E+\Delta E )$ [10]. Putting (4.5) in (4.4)
we obtain the entropy of the system as [10] \begin{center}$$ S = k
\ln \left\{\frac{\Omega (E,V,N)}{C^{N}}\right\}
\eqno(4.6)$$\end{center} The constant $C^{N}$, has the same unit
as $\Omega (E,V,N)$ and cannot be determined classically. However
it can be determined from quantum mechanics. Then we have $C^{N} =
(h)^{3N}$ for distinguishable particles and  $C^{N} = N! (h)^{3N}$
for indistinguishable particles. From Heisenberg uncertainty
principle, we know that if $h$ is the volume of a single state in
phase space then $\Omega (E,V,N)/(h)^{3N}$ is the total number of
microstates in the energy shell $(E, E+\Delta E )$. The expression
(4.6) then becomes identical to the Boltzmann entropy. With this
interpretation of the constant $C^{N}$, the correct expression of
classical entropy is given by [10, 11]
\begin{center}$$S = -k \int_{R} f(X^{N}) \ln \{(h)^{3N} f(X^{N})
\} dX^{N} \eqno(4.7)$$\end{center} The classical entropy that
follows a limiting case of Von Neumann entropy is given by [12]
\begin{center}$$ S_{d} = -k \int_{R} \frac{f(X^{N})}{(h)^{3N}}
\ln \{f(X^{N}) \} dX^{N} \eqno(4.8)$$\end{center} This is,
however different from the one given by (4.7) and it does not
lead to the form of Boltzmann entropy (4.6).

\vspace{2em}

\noindent{\large\bf 6. Conclusion }

\vspace{1em}

The literature on the axiomatic derivation of Shannon entropy is
vast [6, 7]. The present approach is, however, different. This is
based mainly on the postulates of additivity and concavity of
entropy function. There are, infact, variant forms of additivity
and non decreasing characters of entropy in thermodynamics. The
concept of additivity is dormant in many axiomatic derivations of
Shannon entropy. It plays a vital role in the foundation of
Shannon information theory [13]. Non-additive entropies like
Renyi's entropy and Tsallis entropy need a different formulation
and leads to different physical phenomena [14,15]. In the present
paper we have also provided a new axiomatic derivation of Shannon
total entropy which in the limiting case reduces to the expression
of modified differential entropy (4.1). The modified differential
entropy together with quantum uncertainty relation provides a
mathematically strong approach to the derivation of the expression
of classical entropy.

\vspace{3em}

\noindent{\large\bf References }

\vspace{1em}

\begin{enumerate}

\item C. F. Shannon and W. Weaver : {\it Mathematical Theory of
Communication}. University of Illinois Press, Urbana (1949).

\item E. T. Jaynes : Information Theory and Statistical Mechanics.
{\it Phys. Rev.} {\bf 106}(1957), 620-630.

\item G. Jumarie : {\it Relative information and Applications}.
Springer-Verlag, Berlin (1990).

\item J. N. Kapur : {\it Measures of Information and Their
Applications}. Wiley Eastern, New Delhi (1994).

\item V. Majernik : {\it Elementary Theory of Organization}.
Palacky University Press, Olomoue (2001).

\item J. Axzel and Z. Doroc'zy : {\it On Measures of Information
and Their Characterizations}. Academic Press, New York (1975).

\item A. Mathai and R. N. Rathie : {\it Information Theory and
Statistics}. Wiley Eastern, New Delhi (1974).

\item D. Morales, L. pardo and I. Vajda : Uncertainty of Discrete
Stochastic System : General Theory and Statistical Interference.
{\it IEEE Trans. System, Man and Cybernetics A} {\bf 26} (1996),
681-697.

\item T. M. Cover and J. A. Thomas : {\it Elements of Information
Theory}. Wiley and Sons, New York (1991).

\item L. E. Reichl : {\it A Modern Course in Statistical Physics}.
Edwand Arnold (Publ.) Ltd. (1980).

\item L. D. Landau and E. N. Lifshitz : {\it Statistical Physics}.
Pargamon Press, Oxford (1969).

\item A. Wehrl : On the relation between classical entropy and
quantum mechanical entropy {\it Report. Math. Phys.} {\bf 16}
(1979), 353-358.

\item T. Yamano : A Possible Extension of Shannon's Information Theory.
{\it Entropy} {\bf 3} (2001), 280-292.

\item A.Renyi:Probability Theory.North-Holland,Amsterdam(1970)

\item C. Tsallis : Possible Generalization of Boltzmann-Gibbs
Statistics. {\it J. Stat. Phys} {\bf 52} (1988), 479-487.\\

\end{enumerate}
$C.G.Chakraborti:~Department~ of~ Applied~ Mathematics.~
University~ of~
Calcutta.~ Kolkata-700 009, INDIA\\
e-mail :~ cgcappmath@caluniv.ac.in\\
I.Chakrabarty:~Department~ of~ Mathematics.~ Heritage~ Institute~
of~ Technology~ Chowbaga~ Road,~ \\Anandapur~ Kolkata-700 107,~
INDIA
\\e-mail : indranilc@indiainfo.com$
\end{document}